\documentclass[12pt,thmsa]{article}
\usepackage{amsfonts}

\usepackage{amsmath}

\begin{document}

\author{Axel Gelfert$^{(1,2)}$ and Wolfgang Nolting$^{(1)}$ \\
{\footnotesize (1) Humboldt University of Berlin, Department of Physics,}\\
{\footnotesize Invalidenstra\ss e 110, 10115 Berlin, Germany}\\
{\footnotesize and}\\
{\footnotesize (2) University of Cambridge, Department of History and }\\
{\footnotesize Philosophy of Science, Free School Lane, Cambridge CB2 3RH,
U.K.}}
\title{The absence of finite-temperature phase transitions in
low-dimensional many-body models: a survey and new results\thanks{%
Accepted for publication in \textit{Journal of Physics: Condensed Matter.}
\copyright\ Institute of Physics Publishing Ltd. 2001, http://www.iop.org.
For citation refer to printed version only. Please address email to
corresponding author: axel@gelfert.net.}}
\maketitle

\begin{abstract}
After a brief discussion of the Bogoliubov inequality and possible
generalizations thereof, we present a complete review of results concerning
the Mermin-Wagner theorem for various many-body systems, geometries and
order parameters. We extend the method to cover magnetic phase transitions
in the Periodic Anderson Model as well as certain superconducting pairing
mechanisms for Hubbard films. The relevance of the Mermin-Wagner theorem to
approximations in many-body physics is discussed on a conceptual level.
\end{abstract}

\section{Introduction}

The quest for criteria for the existence, or absence, of phase transitions
in physical systems has been a dominant theme in theoretical physics ever
since the phenomenological concept of a \textit{phase} was introduced in the
context of equilibrium thermodynamics which is associated with the names of
Boltzmann, Gibbs, Ehrenfest and many others. The pioneering work in the
phenomenological theory of phase transitions was soon supplemented by
attempts to provide a detailed theoretical description of macroscopic
phenomena via microscopic many-body theories. The Lenz-Ising model (1925)
was the first such model, and Onsager's solution of the 2-D Ising model
(1944) is evidence that, by then, the mathematical physics of phase
transitions had become a subject in its own right. Other landmark
developments of the 1950s and 1960s were the famous Yang and Lee papers
(1952) providing a mathematically rigorous scenario of how phase transitions
can occur in the thermodynamic limit, and the papers by Hohenberg (1967) and
Mermin and Wagner (1966) excluding certain phase transition in
low-dimensional systems.

It is the latter result, generally known as the Mermin-Wagner theorem, which
we shall discuss in this paper. In doing so, our goal is three-fold. First,
we discuss applications, and possible generalizations, of the Bogoliubov
inequality, which underlies the proof of the Mermin-Wagner theorem. Then we
present a survey of existing proofs of the absence of finite-temperature
phase transitions in low-dimensional systems along with recent
generalizations. Finally, by means of a simple example, we discuss the
relevance and validity of the Mermin-Wagner theorem in approximate theories
designed to describe a magnetic phase transition.

\section{Mathematical tools}

The occurrence of a phase transition is often intimately related to the
failure of one of the phases to exhibit a certain symmetry property of the
underlying Hamiltonians. Crystals, for example, by their very lattice
structure, break the translational symmetry encountered in the continuum
description of fluids; ferromagnets, in addition to the spatial
symmetry-breaking due to their crystal structure, are not invariant under
rotations in spin-space, even though the underlying Hamiltonians describing
the system may well be. Less obvious types of symmetry-breaking occur in
other quantum systems, such as superfluids and superconductors, where a
breaking of gauge invariance occurs.

\subsection{Bogoliubov quasi-averages}

Bogoliubov has devised a method for describing the occurrence of spontaneous
symmetry-breaking in terms of\textit{\ quasi-averages.}\cite{Bog 1962}\cite%
{Bog 1960} Normally, for systems in statistical equilibrium, the expectation
value of an operator $A$ is given by the trace over the equilibrium \textit{%
density} (or\textit{\ statistical) operator} $\rho =\exp (-\beta \mathcal{H}%
) $ times the observable $A.$ Thus, for the infinite system $(V\rightarrow
\infty )$ one must calculate 
\begin{eqnarray}
\left\langle A\right\rangle &\equiv &\lim_{V\rightarrow \infty }tr\left(
\rho A\right)  \notag \\
&=&\lim_{V\rightarrow \infty }tr\left( e^{-\beta \mathcal{H}}A\right) ,
\label{averag}
\end{eqnarray}%
where $\mathcal{H}$ is the grand-canonical Hamiltonian, $\mathcal{H}=H-\mu 
\hat{N}$ ($\hat{N}$: number operator). However, it turns out that under
certain conditions such averages may be unstable with respect to an
infinitesimal perturbation of the Hamiltonian. If a corresponding additive
contribution $H_{\nu }\equiv \nu H^{\prime }$ of the order of $\nu $ is
added (where $\nu $ is a small positive number which will eventually be
taken to zero: $\nu \rightarrow 0)$, i.e. 
\begin{equation}
\mathcal{H}_{\nu }=H+H_{\nu }-\mu \hat{N},  \label{calhanu}
\end{equation}%
one can define the \textit{quasi-average} of $A$ in the following way: 
\begin{equation}
\left\langle A\right\rangle _{q}\equiv \lim_{\nu \rightarrow
0}\lim_{V\rightarrow \infty }tr(e^{-\beta \mathcal{H}_{\nu }}A).
\label{quaverag}
\end{equation}%
The average (\ref{averag}) and the quasi-average (\ref{quaverag}) need not
coincide, since the two limits in (\ref{quaverag}) may fail to commute
within some parameter region (i.e. for some combination of $\mu $ and $\beta
)$.

Quasi-averages are appropriate for cases in which spontaneous
symmetry-breaking occurs, as can be shown by a simple argument. Suppose the
Hamiltonian $\mathcal{H}$ displays a continuous symmetry $\mathcal{S}$, i.e.
it commutes with the generators $\Gamma _{\mathcal{S}}^i$ of the
corresponding symmetry group, 
\begin{equation}
\left[ \mathcal{H},\Gamma _{\mathcal{S}}^i\right] _{-}=0.
\end{equation}
If some operator $B$ is not invariant under the transformations of $\mathcal{%
S}$, 
\begin{equation}
\left[ B,\Gamma _{\mathcal{S}}^i\right] _{-}\equiv C^i\neq 0,
\end{equation}
the (normal) average of the commutator $C^i$ vanishes, 
\begin{equation}
\left\langle C^i\right\rangle =0,  \label{vanichi}
\end{equation}
as can be readily seen from eqn. (\ref{averag}) by use of cyclic invariance
of the trace. In those instances, however, where the perturbative part $%
H_\nu $ of (\ref{calhanu}) does not commute with $\Gamma _{\mathcal{S}}^i,$
this will give a \textit{non-vanishing} \textit{quasi-average:} 
\begin{eqnarray}
\left\langle C^i\right\rangle _q &=&\lim_{\nu \rightarrow 0}tr\left(
e^{-\beta \mathcal{H}_\nu }\left[ B,\Gamma _{\mathcal{S}}^i\right]
_{-}\right)  \notag \\
&\neq &0.
\end{eqnarray}
Thus, even though one might na\"{\i}vely expect the quasi-average to
coincide with the ``normal'' average in the limit $\nu \rightarrow 0,$ quite
generally this will not be the case. Note that the quasi-average depends on
the nature of the perturbation added to the ``original'' Hamiltonian.

As a first example, let us examine the Heisenberg model 
\begin{equation}
H=-\sum_{ij}J_{ij}\left( \vec{S}_{i}\cdot \vec{S}_{j}\right) ,
\end{equation}%
which is invariant under the continuous rotation group generated by the
total spin vector $\vec{S}=\sum_{i}\vec{S}_{i}$ because of $\left[ H,\vec{S}%
\right] _{-}=0.$ Thus, we can take $\vec{S}$ as the operator $B$, and from (%
\ref{vanichi}) one finds $\left\langle \left[ S^{\alpha },S^{\beta }\right]
_{-}\right\rangle =0,$ where $S^{\alpha },S^{\beta }$ are the components of
the total spin vector $\vec{S}.$ Together with the commutation relations for
spin operators, e.g. $\left[ S^{x},S^{y}\right] _{-}=i\hbar S^{z},$ it is
obvious that the conventional average of the magnetization vanishes. This is
just a manifestation of the fact, that on the macroscopic level, for an
ideal, infinitely extended system, there is no preferred direction in space.
One can think of this situation as a ``degeneracy'' with respect to spatial
orientation. Adding an external field, $\vec{B}_{0}=B_{0}\vec{e}_{z}$ along
the $z$-axis for example, lifts this degeneracy and via the contribution $%
H_{b}\sim B_{0}M(T,B_{0})$ to the Hamiltonian $(M$: magnetization) one can,
thus, construct appropriate quasi-averages for arbitrary operators according
to equation (\ref{quaverag}).

\subsection{Order parameters}

In the theory of phase transitions, the first step is to identify a quantity
whose (quasi-)average vanishes on one side of the transition, but takes on a
finite value on the other side. This quantity is called the \textit{order
parameter. }In a continuous phase transition, the order parameter may
gradually evolve from zero at the critical point to a finite value on one
side (usually the low-temperature side) of the transition. For different
kinds of phases, different order parameters must be chosen. From a
phenomenological point of view, one must consider each physical system anew.
We have already mentioned the liquid-vapour and the
ferromagnetic-paramagnetic transition as prototypes for phase transitions.
In the former, the obvious choice for the order parameter would be the
difference between the mean densities, i.e. $\rho -\rho _{vapour};$ in the
latter the relevant order parameter is the magnetization $M.$ Within a given
many-body model, the magnetization can be defined in microscopic terms, as
we shall see shortly.

Sometimes, as in the transition to the superconducting state, it may even be
possible to characterize the \textit{same} type of phase transition by use
of different order parameters. According to the standard theories of
superconductivity, at low temperatures electrons with opposite spins form
Cooper pairs. Thus, a possible order parameter would be the average
probability amplitude to find a Cooper pair at a given lattice site in the
crystal. Alternatively, one could characterize the phase transition through
the gap parameter whose modulus is the difference in energy per electron of
the Cooper pair condensate and the energy at the Fermi level. We shall
briefly return to the problem of superconductivity when we consider the
possibility of pairing at finite temperatures in film systems.

\subsection{Bogoliubov inequality}

The Bogoliubov inequality is a rigorous relation between two essentially
arbitrary operators $A$ and $B$ and a valid Hamiltonian $H$ of a physical
system. In its original form, proposed in \cite{Bog 1962}, it is given by 
\begin{equation}
\left| \left\langle \left[ C,A\right] _{-}\right\rangle \right| ^{2}\leq 
\frac{\beta }{2}\left\langle \left[ A,A^{\dagger }\right] _{+}\right\rangle
\left\langle \left[ C^{\dagger },[H,C]_{-}\right] _{-}\right\rangle ,
\label{bogsimple}
\end{equation}%
where $\beta =1/k_{B}T$ is the inverse temperature and $\left\langle
....\right\rangle $ denotes the thermodynamic expectation value. $A$ and $B$
do not necessarily have an obvious physical interpretation from the very
beginning, so the physical significance of (\ref{bogsimple}) will depend on
the suitable choice for the operators involved. The inequality can be proved
by introducing a scalar product which is based on the energy eigenvalues $%
E_{n}$ and the orthogonality of the corresponding energy eigenstates $\left|
n\right\rangle $ of the Hamiltonian $H$ and to which one then applies the
Schwarz inequality. The details of the proof are readily accessible
elsewhere (see e.g. textbook \cite{Nol 1986}); however, we briefly note that
as a result of this derivation, the two factors on the RHS of (\ref%
{bogsimple}) each are \textit{upper bounds to a norm} and can, thus, be
bounded from below by zero. In particular, if, for example, the double
commutator depends on some parameter $k,$ one will always find 
\begin{equation}
\left\langle \left[ \lbrack C,H]_{-},C^{\dagger }\right] _{-}\right\rangle
(k)+\left\langle \left[ [C,H]_{-},C^{\dagger }\right] _{-}\right\rangle
(k^{\prime })\geq \left\langle \left[ \lbrack C,H]_{-},C^{\dagger }\right]
_{-}\right\rangle (k).  \label{sumtransit}
\end{equation}

Dividing both sides of (\ref{bogsimple}) by the double commutator and
summing over all wave vectors $\vec{k}$ associated with the first Brillouin
zone in the reciprocal lattice, one arrives at 
\begin{equation}
\sum_{\vec{k}}\frac{\left| \left\langle \left[ C,A\right] _{-}\right\rangle
\right| ^{2}}{\left\langle \left[ [C,H]_{-},C^{\dagger }\right]
_{-}\right\rangle (\vec{k})}\leq \frac{\beta }{2}\sum_{\vec{k}}\left\langle %
\left[ A,A^{\dagger }\right] _{+}\right\rangle (\vec{k}).  \label{bog}
\end{equation}

\subsection{Generalized Bogoliubov-type inequalities}

The use of inequalities in the theory of phase transitions has developed
into a subdiscipline of its own right within the field of mathematical
physics \cite{Fro Pfi 1981},\cite{Lut 1966}; consequently, this article is
not intended to give a complete account of such approaches. The close
kinship between some of these methods and the Bogoliubov inequality,
however, justifies a brief discussion of what one might call \textit{%
generalized Bogoliubov-type inequalities.} The common feature of these
inequalities is the fact that certain algebraic properties of the quantities
involved allow one to use the Cauchy-Schwarz inequality, thus leading to
upper, or lower, bounds for physical observables, such as the dynamical
structure factor%
\begin{eqnarray}
C_{BA}(\vec{k}^{\prime },\vec{k};E^{\prime }) &=&\int\limits_{-\infty
}^{\infty }dte^{iE^{\prime }t/\hbar }\sum_{n}\frac{e^{-\beta E_{n}}}{N\Xi }%
\left\langle n\left| e^{i\mathcal{H}t/\hbar }B(\vec{k}^{\prime },0)e^{-i%
\mathcal{H}t/\hbar }A(\vec{k},0)\right| n\right\rangle \\
&=&2\pi \hbar \sum_{mn}\frac{e^{-\beta E_{n}}}{N\Xi }\left\langle m\left| A(%
\vec{k},0)\right| n\right\rangle \left\langle n\left| B(\vec{k}^{\prime
},0)\right| m\right\rangle \delta (E^{\prime }-(E_{m}-E_{n}))  \notag
\end{eqnarray}%
(where the sum runs over all energy eigenstates $\left| n\right\rangle ,$ $%
\left| m\right\rangle $ corresponding to eigenvalues $E_{n},$ $E_{m}$), or
the susceptibility%
\begin{equation*}
\chi _{AB}(\vec{k},\vec{k}^{\prime };E,i\eta )=\frac{1}{2\pi }%
\int\limits_{-\infty }^{\infty }dE^{\prime }\frac{\left( 1-e^{-\beta
E^{\prime }}\right) C_{BA}(E^{\prime })}{E-E^{\prime }+i\eta }.
\end{equation*}%
Wagner \cite{Wag 1966} discusses the static susceptibility $\chi _{AB},$
defined as%
\begin{eqnarray}
\chi _{AB}(\vec{k},\vec{k}^{\prime }) &=&\mathfrak{R}\left( \lim_{E,\eta
\rightarrow 0}\chi _{AB}(\vec{k},\vec{k}^{\prime };E,i\eta )\right)  \notag
\\
&=&\mathcal{P}\int\limits_{-\infty }^{\infty }dE^{\prime }\frac{S_{AB}(\vec{k%
},\vec{k}^{\prime };E^{\prime })}{E^{\prime }}  \label{statsusc}
\end{eqnarray}%
in terms of the spectral density which, in its spectral representation, is
given by%
\begin{equation}
S_{AB}(\vec{k},\vec{k}^{\prime };E^{\prime })=\frac{\hbar }{\Xi }%
\sum_{nm}\left\langle n\left| A(\vec{k},0)\right| m\right\rangle
\left\langle m\left| B(\vec{k}^{\prime },0)\right| n\right\rangle e^{-\beta
E_{m}}\left( e^{\beta E^{\prime }}-1\right) \delta (E^{\prime
}-(E_{m}-E_{n})).
\end{equation}

It can then be shown \cite{Wag 1966},\cite{Gel 2001} that 
\begin{equation}
\left\langle A(\vec{k});B(\vec{k}^{\prime })\right\rangle :=\chi _{AB}(\vec{k%
},\vec{k}^{\prime })
\end{equation}%
is a valid scalar product in the space of operators $A,$ $B.$ The Schwarz
inequality for the static susceptibility then reads%
\begin{equation}
\left| \chi _{AB}(\vec{k},\vec{k}^{\prime })\right| ^{2}\leq \chi
_{AA^{\dagger }}(\vec{k})\chi _{B^{\dagger }B}(\vec{k}^{\prime }).
\end{equation}%
Physically significant relations can be deduced from this general statement
by an appropriate choice of operators $A$ and $B.$ If $A$ is taken to be a
time derivative of another operator \cite{Wag 1966}, i.e.%
\begin{equation}
A_{\vec{k}}(t):=i\hbar \frac{\partial }{\partial t}Q_{\vec{k}}(t),
\end{equation}%
the inequality%
\begin{equation}
\mathcal{P}\int\limits_{-\infty }^{\infty }dE\frac{S_{B^{\dagger }B}(\vec{k}%
;E)}{E}\geq \frac{\left| \left\langle \left[ Q_{\vec{k}},B_{\vec{k}^{\prime
}}\right] _{-}\right\rangle \right| ^{2}}{\left\langle \left[ \left[ Q_{\vec{%
k}},H\right] _{-},Q_{\vec{k}}^{\dagger }\right] _{-}\right\rangle }
\label{suscin}
\end{equation}%
follows, which relates the response function for the observable $B$ to
commutators that can, in principle, be calculated directly from one's
knowledge of $Q,$ $B,$ and $H.$ If for a given many-body Hamiltonian $H$ one
specifies $Q$ and $B$ further, it is possible to obtain bounds for
non-trivial order parameters. Thus, for the planar magnetization $M_{p}$
defined as%
\begin{equation}
M_{p}(T,\tilde{B}_{0}):=\frac{1}{N}\sum_{i}\left\langle \alpha _{i}\sigma
_{i}^{y}+\beta _{i}\sigma _{i}^{x}\right\rangle
\end{equation}%
(where $i$ runs over all lattice sites, $\sigma _{i}^{x}$ and $\sigma
_{i}^{y}$ denote the respective components of the spin, and the real
constants $\alpha _{i}$ and $\beta _{i}$ can be chosen so as to take into
account various kinds of spin ordering within the $xy$-plane), evaluating (%
\ref{suscin}) with operators $Q$ and $B$ given by%
\begin{equation*}
Q_{\vec{k}}=\sum_{j}e^{-i\vec{k}\cdot \vec{R}_{j}}\left( \beta _{j}\sigma
_{j}^{y}-\alpha _{j}\sigma _{j}^{x}\right)
\end{equation*}%
and%
\begin{equation*}
B_{\vec{k}}=\sum_{j}e^{-i\vec{k}\cdot \vec{R}_{j}}\sigma _{j}^{z}=\sigma
^{z}(\vec{k})
\end{equation*}%
gives an upper bound for $M_{p}$ in terms of the longitudinal susceptibility 
$\chi ^{zz}(\vec{k}).$ Since for the latter, rigorous limits and estimates
have been established, e.g. for the attractive Hubbard model \cite{Kub Kis
1990}, these carry over to the order parameter $M_{p}.$

Before turning to exact results concerning finite-temperature phase
transitions, let us discuss how correlation inequalities can be put to use
in the zero-temperature case. Pitaevskii and Stringari \cite{Pit Str 1991}
have suggested to define a scalar product simply through the anticommutator: 
\begin{equation}
\left\langle A;B\right\rangle :=\left\langle [A^{\dagger
},B]_{+}\right\rangle
\end{equation}%
which gives the Schwarz inequality 
\begin{equation}
\left\langle \lbrack A^{\dagger },A]_{+}\right\rangle \left\langle
[B^{\dagger },B]_{+}\right\rangle \geq \left| \left\langle \lbrack
A^{\dagger },B]_{+}\right\rangle \right| ^{2}.  \label{schwarzpitstring}
\end{equation}%
It is then possible to define auxiliary operators, denoted by a tilde, such
that 
\begin{eqnarray}
&\left\langle n\left| \tilde{C}\right| 0\right\rangle :=&\left\langle
n\left| C\right| 0\right\rangle  \notag \\
&\left\langle 0\left| \tilde{C}\right| n\right\rangle :=&-\left\langle
0\left| C\right| n\right\rangle
\end{eqnarray}%
from which it follows that 
\begin{eqnarray}
\left\langle n\left| \tilde{C}^{\dagger }\right| 0\right\rangle
&=&-\left\langle n\left| C^{\dagger }\right| 0\right\rangle  \notag \\
\left\langle 0\left| \tilde{C}^{\dagger }\right| n\right\rangle
&=&\left\langle 0\left| C^{\dagger }\right| n\right\rangle .
\end{eqnarray}%
At $T=0$ the expectation values in (\ref{schwarzpitstring}) are evaluated in
the ground state $\left| 0\right\rangle ,$ and if instead of $B$ the newly
defined operator $\tilde{B}$ is used, one deduces from 
\begin{eqnarray}
\left\langle 0\left| [A^{\dagger },\tilde{B}]_{+}\right| 0\right\rangle
&=&\left\langle 0\left| [A^{\dagger },B]_{-}\right| 0\right\rangle  \notag \\
\left\langle 0\left| [\tilde{B}^{\dagger },\tilde{B}]_{+}\right|
0\right\rangle &=&\left\langle 0\left| [B^{\dagger },B]_{+}\right|
0\right\rangle  \label{elimtilde}
\end{eqnarray}%
that the Schwarz inequality holds not only for the anticommutator on the RHS
but also for the commutator, so that (writing down (\ref{schwarzpitstring})
for the operators $A,\tilde{B},$ and making use of (\ref{elimtilde}) to
express $\tilde{B}$ in terms of $B)$ one finds 
\begin{equation}
\left\langle \lbrack A^{\dagger },A]_{+}\right\rangle \left\langle
[B^{\dagger },B]_{+}\right\rangle \geq \left| \left\langle \lbrack
A^{\dagger },B]_{-}\right\rangle \right| ^{2}.  \label{commuschwarz}
\end{equation}%
With this in mind, one can turn to the dynamical structure factor, which for
the transverse spin-spin correlation function at $T=0$ is given by 
\begin{eqnarray}
\left. C^{+-}(\vec{q},E)\right| _{T=0} &\equiv &\frac{1}{N}%
\int\limits_{-\infty }^{\infty }dt\left. e^{iEt/\hbar }\left\langle \sigma
^{+}(\vec{q},t)\sigma ^{-}(-\vec{q},0)\right\rangle \right| _{T=0}  \notag \\
&=&\frac{1}{N}\int\limits_{-\infty }^{\infty }dte^{iEt/\hbar
}\sum_{m}\left\langle 0\left| e^{i\mathcal{H}t/\hbar }\sigma ^{+}(\vec{q}%
,0)e^{-i\mathcal{H}t/\hbar }\right| m\right\rangle \left\langle m\left|
\sigma ^{-}(-\vec{q},0)\right| 0\right\rangle  \notag \\
&=&\frac{2\pi \hbar }{N}\sum_{m}\delta (E-(E_{m}-E_{0}))\left| \left\langle
0\left| \sigma ^{+}(\vec{q},0)\right| m\right\rangle \right| ^{2}.
\end{eqnarray}%
Since $E_{0}$ is the ground state energy, we have $E_{m}\geq E_{0}$, so the $%
\delta $-function only gives a contribution if $E\geq 0.$ As a corollary 
\begin{equation}
EC^{+-}(\vec{q},E)\left\{ 
\begin{array}{l}
\geq 0 \\ 
=0%
\end{array}%
\right. \text{if}%
\begin{array}{l}
E>0 \\ 
E\leq 0.%
\end{array}
\label{positivity}
\end{equation}

If one defines 
\begin{eqnarray}
C_{\perp }(\vec{q},E) &=&\frac 1{2}\left( C^{+-}(\vec{q},E)+C^{-+}(-\vec{q}%
,E)\right)  \notag \\
&=&\frac 1{2}\left\langle \left[ \sigma ^{+}(\vec{q},E),\sigma ^{-}(-\vec{q}%
,E)\right] _{+}\right\rangle
\end{eqnarray}
and 
\begin{equation}
C_{\perp }(\vec{q})=\int dEC_{\perp }(\vec{q},E),
\end{equation}
one can make use of (\ref{commuschwarz}) for operators $\sigma ^{+}(\vec{q}+%
\vec{Q}),$ $\sigma ^{-}(-\vec{q})$ to get 
\begin{equation}
C_{\perp }(\vec{q}+\vec{Q})C_{\perp }(\vec{q})\geq \left| M(\vec{Q})\right|
^2,
\end{equation}
where the magnetization 
\begin{equation}
M(\vec{Q})=\frac{2\hbar }N\left\langle \sum_i\sigma _i^ze^{-i\vec{Q}\cdot 
\vec{R}_i}\right\rangle
\end{equation}
has been introduced (the reciprocal lattice vector $\vec{Q}$ as usual
accounts for possible antiferromagnetic order). Eqn. (\ref{positivity})
tells us that $C_{\perp }(\vec{q},E)$ is nonvanishing only for $E>0$, where
it is positive, so the integral quantity $C_{\perp }(\vec{q})$ may be
formally written as 
\begin{equation}
C_{\perp }(\vec{q})=\int dE\sqrt{EC_{\perp }(\vec{q},E)}\cdot \sqrt{\frac{%
C_{\perp }(\vec{q},E)}E}.  \label{hoeldereins}
\end{equation}

In this form, one can apply \textit{H\"{o}lder's inequality} which states
that for real-valued function $f$ and $g$ for which $\left| f(x)\right| ^{p}$
and $\left| g(x)\right| ^{q}$ are integrable (where $p$ and $q$ are numbers
satisfying the relations $p^{-1}+q^{-1}=1$ and $p>1)$ the following
inequality holds: 
\begin{equation}
\int_{a}^{b}dxf(x)g(x)\leq \left( \int_{a}^{b}dx\left| f(x)\right|
^{p}\right) ^{1/p}\left( \int_{a}^{b}dx\left| g(x)\right| ^{q}\right) ^{1/q}.
\label{hoeldmath}
\end{equation}

For $p=q=2$ and with $f,g$ defined as 
\begin{eqnarray}
f(x) &=&\sqrt{EC_{\perp }(\vec{q},E)}  \notag \\
g(x) &=&\sqrt{\frac{C_{\perp }(\vec{q},E)}{E}}
\end{eqnarray}%
H\"{o}lder's inequality (\ref{hoeldmath}) applied to (\ref{hoeldereins})
gives 
\begin{equation}
C_{\perp }(\vec{q})\leq \sqrt{\int dEEC_{\perp }(\vec{q},E)}\cdot \sqrt{\int
dE\frac{C_{\perp }(\vec{q},E)}{E}}.  \label{hoelderzwei}
\end{equation}

The first of the integrals can in many cases be bounded from above by 
\begin{eqnarray}
\int dEEC_{\perp }(\vec{q},E) &=&\left\langle \left[ \sigma ^{+}(\vec{q},E),%
\left[ \mathcal{H},\sigma ^{-}(-\vec{q},E)\right] _{-}\right]
_{-}\right\rangle  \notag \\
&\equiv &f(\vec{q})Nq^{2}\leq const\cdot Nq^{2}
\end{eqnarray}%
which suggests writing 
\begin{equation}
\frac{\int dEEC_{\perp }(\vec{q},E)}{\int dE\frac{C_{\perp }(\vec{q},E)}{E}}%
\equiv \omega ^{2}(\vec{q})\cdot q^{2},  \label{omega}
\end{equation}%
thus defining the new quantity $\omega (\vec{q}).$ Inequality (\ref%
{hoeldereins}) can be strengthened using (\ref{hoelderzwei}) and rearranged
by use of (\ref{omega}) to give 
\begin{equation}
C_{\perp }(\vec{q}+\vec{Q})\geq \frac{\left| M(\vec{Q})\right| ^{2}}{\omega (%
\vec{q})q\cdot \chi ^{+-}(\vec{q})}.  \label{strucfacscale}
\end{equation}

This equation relates the order parameter $M,$ the transverse susceptibility 
$\chi ^{+-}$ and the structure factor $C^{+-}$ (via $C_{\perp })$. The
relation between these quantities can be made even more pronounced by
summing both sides of (\ref{strucfacscale}) over the first Brillouin zone.
One can then go on to show that in the thermodynamic limit a relation of the
form 
\begin{equation}
\xi _{0}\geq \frac{v^{(d)}}{(2\pi )^{d}}\int\limits_{1.B.Z.}d^{d}\vec{q}%
\frac{1}{q}\frac{\left| M(\vec{Q})\right| ^{2}}{\omega (\vec{q})\chi ^{+-}(%
\vec{q})}.  \label{xinull}
\end{equation}%
holds (where $\xi _{0}$ turns out to be a model-dependent constant).
Provided that $\omega (\vec{q})$ and $\chi ^{+-}(\vec{q})$ remain finite in
the limit $q\rightarrow 0,$ this means that in 1D not even at zero
temperature can there be a transition to a state with nonvanishing $M(\vec{Q}%
)$, as this would lead to a logarithmic divergence of the RHS and thus would
contradict $\xi _{0}$ being a constant.

It seems to us that the range of applications of correlation inequalities
has not yet been fully exhausted. Using the mathematical tools discussed in
the previous paragraphs, and variations thereof, should serve as a starting
point for the derivation of similar relations for other order parameters and
many-body models.

\section{The Mermin-Wagner theorem: A survey}

Having established the wide range of relationships the Bogoliubov, and
related, inequalities can generate, we shall now turn to a narrower range of
applications, namely results concerning the absence of finite-temperature
phase transitions in low-dimensional systems.

Hohenberg \cite{Hoh 1967} was the first to note that the Bogoliubov
inequality could be used to exclude phase transitions, showing that there
could be no finite-temperature phase transition in one- and two-dimensional
superfluid systems. At roughly the same time Mermin and Wagner, following a
suggestion by Hohenberg, considered the case of spontaneous magnetization in
the Heisenberg model.\cite{Mer Wag 1966} Since Mermin and Wagner's proof has
become the exemplar for studies concerning the absence of phase transitions,
we shall briefly outline the paradigmatic procedure (see also \cite{Gel Nol
2000}). The general idea is to use the Bogoliubov inequality (\ref{bog}) to
find an upper bound $f(B_{0},M)$ for the order parameter in question, e.g.
the spontaneous magnetization $M:$%
\begin{equation}
M\leq f(B_{0},M).  \label{mleqf}
\end{equation}%
As indicated, the upper bound will normally depend on the external (e.g.
magnetic) field that couples to the order parameter, and (implicitly) on the
order parameter itself. To answer the question whether or not a phase
transition to a state with a non-zero value of the order parameter occurs,
one must consider the case $B_{0}\rightarrow 0$, i.e. the behaviour of the
upper bound in the case of vanishing external field. The subsequent argument
against a phase transition proceeds by \textit{reductio ad absurdum: }If the
assumption $M\neq 0$ can be shown to lead to a violation of (\ref{mleqf}) in
the limit $B_{0}\rightarrow 0$, where equation (\ref{mleqf}) is derived from
the Bogoliubov equation (which is known to hold for the corresponding
many-body system), it must be dropped. This, then, leaves as the only
conclusion that $M_{\overrightarrow{B_{0}\rightarrow 0}}0$, the case of
vanishing order parameter and no phase transition. This argument, of course,
only succeeds if the initial inequality is indeed true \textit{``a priori,''}
and the most straightforward way to achieve this is to resort to the
Bogoliubov inequality. From this it follows that, once one has specified a
many-body model by its Hamiltonian $H,$ the operators $A$ and $C$ in the
Bogoliubov inequality must be carefully chosen so as to give the desired
order parameter.

As an example, consider the Heisenberg model%
\begin{equation}
H_{(Hei)}=-\sum_{ij=1...N}J_{ij}\left(
S_{i}^{+}S_{j}^{-}+S_{i}^{z}S_{j}^{z}\right) -b\sum_{i=1...N}e^{-i\vec{K}%
\cdot \vec{R}_{i}}S_{i}^{z}
\end{equation}%
(where the term $b\sum_{i}e^{-i\vec{K}\cdot \vec{R}_{i}}S_{i}^{z}$ is due to
the interaction with an external magnetic field $b=\frac{g_{J}\mu _{B}B_{0}}{%
\hbar }).$ The relevant order parameter is the magnetization 
\begin{equation}
M=\frac{1}{N}\frac{g_{J}\mu _{B}}{\hbar }\sum_{i}e^{-i\vec{K}\cdot \vec{R}%
_{i}}\left\langle S_{i}^{z}\right\rangle  \label{magnetiz}
\end{equation}%
where the factor $e^{-i\vec{K}\cdot \vec{R}_{i}}$ already accounts for
antiferromagnetic order by changing the sign of the spins in one sublattice
(given that $\vec{K}$ has been properly chosen to achieve just that). This
suggests that an upper bound for $\left\langle S^{z}\right\rangle $ is
essential. Comparison with (\ref{bog}) suggests the choice%
\begin{eqnarray}
A &=&S^{-}(-\vec{k}-\vec{K}) \\
C &=&S^{+}(\vec{k})
\end{eqnarray}%
as, by virtue of the commutation relations, 
\begin{equation}
\left[ S^{+}(\vec{k}_{1}),S^{-}(\vec{k}_{2})\right] _{-}=2\hbar S_{\alpha
}^{z}(\vec{k}_{1}+\vec{k}_{2}),
\end{equation}%
this will indeed give a contribution proportional to the magnetization on
the left-hand side of the Bogoliubov inequality: 
\begin{equation}
\left\langle \left[ C,A\right] _{-}\right\rangle =\xi _{1}NM(T,B_{0}).
\end{equation}%
($\xi _{(i)}$ denote constants depending, at most, on fixed parameters of
the many-body model, i.e. in this case the Heisenberg exchange integrals.)
The other (anti-)commutators that feature in the Bogoliubov inequality can
be bounded from above, as has been shown by Mermin and Wagner:%
\begin{eqnarray}
\left\langle \left[ \left[ C,H\right] _{-},C^{\dagger }\right]
_{-}\right\rangle &\leq &\xi _{2}^{2}N\left( \left| B_{0}M(T,B_{0})\right|
+\xi _{3}\vec{k}^{2}\right) \\
\sum_{\vec{k}}\left\langle \left[ A,A^{\dagger }\right] _{+}\right\rangle
&\leq &2\xi _{4}N^{2}.
\end{eqnarray}%
For the Bogoliubov inequality, 
\begin{equation}
\sum_{\vec{k}}\frac{\left| \left\langle \left[ C,A\right] _{-}\right\rangle
\right| ^{2}}{\left\langle \left[ [C,H]_{-},C^{\dagger }\right]
_{-}\right\rangle }\leq \frac{\beta }{2}\sum_{\vec{k}}\left\langle \left[
A,A^{\dagger }\right] _{+}\right\rangle ,
\end{equation}%
one finds in this case 
\begin{equation*}
\sum_{\vec{k}}\frac{\xi _{1}^{2}N^{2}M^{2}(T,B_{0})}{\xi _{2}^{2}N\left(
\left| B_{0}M(T,B_{0})\right| +\xi _{3}\vec{k}^{2}\right) }\leq \xi
_{4}\beta N^{2}.
\end{equation*}%
In the thermodynamic limit, the only situation where one can at all hope for
a phase transition, the sum is to be replaced by an integral, e.g. for the
two-dimensional case%
\begin{equation}
\sum_{\vec{k}}\hat{=}\frac{L^{2}}{(2\pi )^{2}}\int\limits_{\vec{k}}d^{2}\vec{%
k},  \label{transitiontotdl}
\end{equation}%
where $L^{2}/(2\pi )^{2}$ is the area in $k$-space associated with one
quantum state. Restricting the support of the integral to a finite-volume
sphere inscribed into the first Brillouin zone only strengthens the
inequality, so 
\begin{equation}
\left( \frac{\xi _{1}}{\xi _{2}}\right) ^{2}\frac{1}{2\pi }\frac{L^{2}}{N}%
M^{2}(T,B_{0})\int\limits_{0}^{k_{0}}dk\frac{k}{\left|
B_{0}M(T,B_{0})\right| +\xi _{3}k^{2}}\leq \xi _{4}\beta
\end{equation}%
where $k_{0}$ is the cutoff corresponding to the sphere in $k$-space. In the
thermodynamic limit under consideration, $L$ and $N$ approach infinity in
such a way that the specific volume $v_{0}^{(2)}=L^{2}/N$ remains finite
throughout. Evaluating the integral, one then arrives at%
\begin{equation}
M^{2}(T,B_{0})\leq \xi \frac{\beta }{\ln \left( 1+\frac{\xi _{3}k_{0}^{2}}{%
\left| B_{0}M(T,B_{0})\right| }\right) }.  \label{magup}
\end{equation}%
As $B_{0}\rightarrow 0,$ the denominator diverges logarithmically, thus
forcing the magnetization to vanish. This result is independent of the
original choice of the auxiliary wave vector $\vec{K}$ (see (\ref{magnetiz}%
)), so that both ferromagnetic and antiferromagnetic order is ruled out in
the two-dimensional Heisenberg model. One easily verifies that a similar
divergence of the denominator rules out spontaneous magnetic order in the
one-dimensional case.

The main steps in any proof of the Mermin-Wagner type are: a) choice of a
many-body model characterized by its Hamiltonian $H;$ b) identification of
the order parameter for the phase transition to be discussed; c) adequate
choice of operators $A$ and $B$ in the Bogoliubov inequality so as to single
out the order parameter; d) search for non-trivial upper bounds for the
(anti-)commutators in the Bogoliubov inequality; e) proof that in the
thermodynamic limit the assumption of a spontaneous non-zero value for the
order parameter will be self-refuting in the one- or two-dimensional case
(and, possibly, other cases as well). This basic scheme has been applied to
a wide range of different models and order parameters. The remainder of this
section attempts to present a survey of established as well as new results,
along with brief discussions and related references.

Shortly after the papers by Hohenberg \cite{Hoh 1967} and Mermin and Wagner %
\cite{Mer Wag 1966}, Mermin showed that a classical Bogoliubov-type
inequality holds, if in equation (\ref{bogsimple}) one replaces the
commutators by Poisson brackets and also requires certain quantities in the
corresponding classical thermal averages to vanish. For short-range
interactions, one can then rule out phase transitions for the ``classical,''
i.e. infinite-spin limit.\cite{Mer 1967}

Walker used the Bogoliubov inequality to rule out the possibility of a phase
transition, in one and two dimensions, to an excitonic insulating state.\cite%
{Wal 1967} The main idea behind the formation of the insulating state
involved the assumption that certain conditions might favour bound
electron-hole pairs to form if a semiconductor with a very small gap, or a
semimetal with a very small band overlap, was to be cooled to a sufficiently
low temperature. The proof that such a state is impossible in the
low-dimensional case is straightforward for electrons interacting via a
potential which falls off faster than $\left| \vec{r}\right| ^{-D}$ ($D$:
number of dimensions); it can be generalized to cover certain simplified
Hamiltonians with weaker convergence behaviour. A generalization to an
isotropic two-band model can also be given.\cite{Wal 1967}

The Bogoliubov inequality, together with an analogous classical inequality,
also rules out the possibility of crystalline ordering in two dimensions,
thus confirming earlier suggestions by Peierls and Landau that there could
be no two-dimensional crystalline ordered state.\cite{Mer 1968} This is
significant, since the extension to crystalline ordering is not quite
straightforward: Contrary to types of ordering where the energies of
fluctuations that cause the disorder are kinetic (e.g. superfluid,
superconducting, or excitonic interactions), the relevant energy in the
crystalline case is potential. Also, one cannot simply posit that one
particle only interacts with a finite number of neighbours (as one can in
the case of spin systems; see next section), because a given particle may
well diffuse through the crystal and interact with many, and possibly all,
others. The proof can nevertheless be achieved by assuming a system of
identical particles with pair potential $\Phi (\vec{r}),$ where $\Phi $ as
well as the related function%
\begin{equation}
\Psi (\vec{r})=\Phi (\vec{r})-\lambda \left| \vec{r}\right| ^{2}\left|
\Delta \Phi (\vec{r})\right|
\end{equation}%
are required to satisfy the criteria for the existence of a proper
thermodynamic limit for a sufficiently small positive value of $\lambda $.%
\cite{Gri 1972} While this includes Lennard-Jones type potentials, it does
not cover hard-core potentials.

The corresponding result for the quantum case, only sketched in \cite{Mer
1968} in an appendix, has been reproduced in detail by Fern\'{a}ndez.\cite%
{Fer 1970b} The paper discusses an electrically neutral sytem of nuclei and
electrons which is confined to one or two finite dimension(s) with the
remaining two or one dimension(s) being infinite. The paradigmatic examples
for such geometrically restricted systems are the infinite slab and the rod
with rectangular cross section and infinite length (see section 3.2 below
for further examples of this kind). One then proceeds to show that for
systems of this kind, no \textit{maximum} long-range crystalline order can
exist, or, formally:%
\begin{equation}
\left\langle \Psi _{\vec{K}}\right\rangle =0\text{ for any }\vec{K}\neq 0,
\end{equation}%
where 
\begin{eqnarray}
\left\langle \Psi _{\vec{K}}\right\rangle &=&\lim_{N\rightarrow \infty
}\left\langle \rho _{\vec{K}}\right\rangle /N \\
\text{with }\rho _{\vec{K}} &=&\int d\vec{r}\rho (\vec{r})\exp \left( -i\vec{%
K}\cdot \vec{r}\right)
\end{eqnarray}%
and $\rho (\vec{r})$ is the number-density operator, and the limit is taken
with $N/V$ being held constant $(V$: volume of the system).

Kishore and Sherrington \cite{Kis She 1972} considered a quite general
Hamiltonian of electrons interacting non-relativistically among themselves,
and with spatially ordered or disordered scatterers, excluding spontaneous
low-dimensional magnetic order. The restriction to non-relativistic
interactions means that spin-orbit effects are excluded from the problem,
thus not ruling out a phase transition for Ising-type, or other suitable
anisotropies. The role of disorder and impurities has also been discussed by
Schuster,\cite{Sch 1977} who discusses the influence of a static random
field conjugate to the order parameter. Such random fields might be
generated by electrically charged or magnetic impurities in certain quenched
systems. The Bogoliubov inequality, in that case, allows one to exclude
magnetic order for a classical $X-Y$ model with Gaussian distributed random
field in less than or equal to four dimensions. The proof makes use of
Mermin's classical analogue of the Bogoliubov inequality mentioned earlier
in this section.\cite{Mer 1967} The significance of the Mermin-Wagner
theorem for the $X-Y$ model and its characteristic
Kosterlitz-Thouless-Berezinskii transition is discussed in \cite{Gul Gul
1996}. For the $X-Y$ model it has also been pointed out \cite{Bra Hol 1994}
that finite-size magnetization may obscure Mermin-Wagner-type behaviour for
any realizable physical system.

\subsection{Spin systems}

Much research has been carried out in order to extend the statement to spin
systems other than the simple Heisenberg model considered by Mermin and
Wagner. Wegner \cite{Weg 1967} considered a model describing a system with
locally interacting itinerant electrons with pseudospins $\vec{\sigma},$ to
which the $B_{0}$-field couples, giving a contribution $\sim B_{0}\sum_{\vec{%
k}}\left( a_{\vec{k}\uparrow }^{\dagger }a_{\vec{k}\uparrow }-a_{\vec{k}%
\downarrow }^{\dagger }a_{\vec{k}\downarrow }\right) $ to the Hamiltonian.
The dynamics of the electrons, in this model, are governed solely by the
kinetic energy $T_{e}=\sum_{\vec{k}\sigma }\frac{k^{2}}{2m}a_{\vec{k}\sigma
}^{\dagger }a_{\vec{k}\sigma }.$ The interaction $V$ of the electrons with
the nuclei and among themselves, is further assumed to satisfy the condition 
\begin{equation}
\left[ C,V\right] _{-}=0  \label{wegnercrit}
\end{equation}%
(where $C$ is the operator that appears in the Bogoliubov inequality).

A more realistic model has been discussed by Walker and Ruijgrok \cite{Wal
Rui 1968}: It includes Coulomb and exchange effects, with possible non-local
interaction. The authors consider a model for which Lieb and Mattis had
previously ruled out ferromagnetic ordering in one dimension \cite{Lie Mat
1962}, and which turns out to be a special case of their general many-band
model for interacting electrons in a metal. Ghosh \cite{Gho 1971}, more
specifically, recovered the Mermin-Wagner theorem for the Hubbard model
given by%
\begin{equation}
H_{Hub}=\sum_{ij}\sum_{\sigma }T_{ij}c_{i\sigma }^{\dagger }c_{j\sigma
}+U\sum_{i}n_{i\sigma }n_{i-\sigma }+\tilde{B}_{0}\sum_{l}(n_{l\uparrow
}-n_{l\downarrow })\exp (-i\vec{K}\cdot \vec{R}_{l}).
\end{equation}

Van den Bergh and Vertogen \cite{Ber Ver 1974} present a proof of the
Mermin-Wagner model for the s-d interaction model characterized by the
Hamiltonian%
\begin{equation}
H_{s-d}=\sum_{lm\sigma }T_{lm}c_{l\sigma }^{\dagger }c_{m\sigma
}-\sum_{lm}J_{lm}\left( \left( c_{l\uparrow }^{\dagger }c_{l\uparrow
}-c_{l\downarrow }^{\dagger }c_{l\downarrow }\right) S_{m}^{z}+c_{l\uparrow
}^{\dagger }c_{l\downarrow }S_{m}^{-}+c_{l\downarrow }^{\dagger
}c_{l\uparrow }S_{m}^{+}\right) .
\end{equation}%
Just as in the case discussed by Walker and Ruijgrok, the s-d interaction
does not satisfy Wegner's criterion (\ref{wegnercrit}). Furthermore, the
system is composed of two subsystems, so that one has to show that both the
conduction electron system and the system of localized magnetic moments fail
to exibit a magnetic phase transition in 1D and 2D.

Robaszkiewicz and Micnas have generalized the proof for the s-d model by
including interactions within the spins of each subsystem and Hubbard-type
interactions; the model, thus, covers the modified Zener model, the extended
Hubbard model and s-d models as particular cases.\cite{Rob Mic 1976}

The range of validity of the Mermin-Wagner model can, of course, also be
extended by discussing more general geometries: Baryakhtar and Yablonskii %
\cite{Bar Yab 1975}, for example, have shown that the Mermin-Wagner theorem
remains valid for systems with an arbitrary number of magnetic sublattices,
and also excludes non-collinear magnetic order when an external field is
applied. Thorpe \cite{Tho 1971} considers the case of ferromagnetism in
phenomenological models with double and higher-order exchange terms; similar
results were obtained in the multi-sublattice case by Krzemi\'{n}ski.\cite%
{Krz 1976} These methods differ from others in that the Hamiltonian is
written as a series expansion in terms of spin spherical harmonics, thus
offering a more systematic way of evaluating the Bogoliubov inequality using
the defining properties of spin spherical harmonics. A closely related
proof, using spherical tensor operators, has been put forward for the
problem of ordering in quadrupolar systems of restricted dimensionality.\cite%
{Rit Mav 1978}

Uhrig \cite{Uhr 1992} has shown that, at finite temperatures, there can be
no planar magnetic order in the one- and two-dimensional generalized
multiband Hubbard model%
\begin{equation}
H_{gen.Hub}=-\sum_{\substack{ ij\sigma  \\ \alpha \gamma }}T_{i\alpha
,j\gamma }c_{i\alpha \sigma }^{\dagger }c_{j\gamma \sigma }+\sum_{\substack{ %
ij  \\ \alpha \gamma }}U_{i\alpha ,j\gamma }n_{i\alpha }n_{j\gamma }.
\end{equation}%
Planar magnetic order, in this case, is taken to be characterized by an
order parameter of the form%
\begin{equation}
M_{plan}=\sum_{i\alpha }\left( \eta _{i\alpha }\sigma _{i\alpha }^{x}+\zeta
_{i\alpha }\sigma _{i\alpha }^{z}\right) ,
\end{equation}%
where $\sigma _{i\alpha }^{(x,y,z)}$ denote the pseudo-spin components at
lattice site $i$, $\alpha $ is the band index, and $\eta _{i\alpha },$ $%
\zeta _{i\alpha }$ are essentially arbitrary real constants that fix the
direction and the norm of the order parameter $M_{plan}.$ Except for the
choice of operators $C$ and $A$ in the Bogoliubov inequality, the proof of
the Mermin-Wagner theorem carries through in the usual way.

Attempts to generalize the Mermin-Wagner theorem to anisotropic models with $%
n$-th nearest neighbour exchange interactions produce results that seem
somewhat artificial.\cite{Mat Mat 1997} Similarly, attempts to extend the
Mermin-Wagner theorem to special cases in three dimensions remain somewhat
incon\-clusive.\cite{Ras Tas 1989} In both cases, a certain circularity
enters the argument: the parameters of the respective models are chosen 
\textit{``ex post facto''} so as to enforce the absence of a phase
transition.

A comparatively large class of many-body models has been discussed by
Proetto and Lopez.\cite{Pro Lop 1983} They confirm the Mermin-Wagner theorem
for Anderson and Kondo lattices, which is of interest since the nature of
the exchange interaction e.g. in the Anderson model is quite different from
the Heisenberg model and variants thereof. The effective exchange
interactions are higher-order functions of the hybridization matrix elements
and are mediated by the conduction band. As can be shown via a canonical
transformation, at higher order multisite interactions exist that go beyond
the simpler models discussed thus far. The full Hamiltonian for this case is
given by%
\begin{eqnarray}
H_{And} &=&\sum_{ij\sigma }t_{ij}c_{i\sigma }^{\dagger }c_{j\sigma
}+\varepsilon _{f}\sum_{i\sigma }f_{i\sigma }^{\dagger }f_{i\sigma }+\frac{U%
}{2}\sum_{i\sigma }f_{i\sigma }^{\dagger }f_{i\sigma }f_{i-\sigma }^{\dagger
}f_{i-\sigma }+\frac{G}{2}\sum_{i\sigma }c_{i\sigma }^{\dagger }c_{i\sigma
}c_{i-\sigma }^{\dagger }c_{i-\sigma }  \notag \\
&&+J\sum_{i}\left( \sum_{\sigma }f_{i\sigma }^{+}f_{i\sigma }\right) \left(
\sum_{\sigma ^{\prime }}c_{i\sigma ^{\prime }}^{\dagger }c_{i\sigma ^{\prime
}}\right) +\sum_{ij\sigma }\left( V(\vec{R}_{i}-\vec{R}_{j})c_{i\sigma
}^{\dagger }f_{i\sigma }+h.c.\right)  \notag \\
&&-\frac{\tilde{B}_{0}}{2}\sum_{l}\left( (f_{l\uparrow }^{\dagger
}f_{l\uparrow }-f_{l\downarrow }^{\dagger }f_{l\downarrow })+(c_{i\uparrow
}^{\dagger }c_{i\uparrow }-c_{i\downarrow }^{\dagger }c_{i\downarrow
})\right) \exp \left( -i\vec{K}\cdot \vec{R}_{l}\right) ,
\end{eqnarray}%
thus covering Hubbard model, Anderson model, Falikov-Kimball model, and
Kondo-lattice model as special cases. The Anderson model has also been
discussed by Noce and Cuoco who rederive the Mermin-Wagner theorem for the
magnetic phase transition. Their paper \cite{Noc Cuo 1999} also presents
analogous proofs for different pairing mechanisms and the corresponding
superconducting phase transitions. For the Hubbard model, a proof of the
absence of superconducting long-range order for different kinds of pairing
has been given in ref.\cite{Su Sch Zit 1997}. A generalization of these
cases will be discussed in section 3.3 below, along with its implications
for superconductivity in thin films.

\subsection{Partially restricted systems}

Early after the publication of the papers by Hohenberg, Mermin, and Wagner,
some authors had expressed their doubt as to the applicability of the
Mermin-Wagner scheme for sytems of finite cross section (or finite
thickness). As we have seen in the previous section, the scheme based on
Bogoliubov's inequality ultimately rests on the divergence of the integral%
\begin{equation}
\int_{k\leq k_{0}}\frac{d^{D}\vec{k}}{k^{2}}
\end{equation}%
in one or two dimensions $(D=1,2).$ The doubts were founded on the
observation that the wave function of a system in a box of finite cross
section must vanish on the walls, thus forcing the wave function to have
some non-zero curvature even in the ground state. This is not compatible
with the wave function assuming a uniform value -- which, it was believed,
should correspond to the $\vec{k}=0$ value (``vanishing momentum'') that
gives the divergence in the first place. Chester et al. \cite{Che Fis Mer
1969} have argued that these doubts are ill-founded, since the $\vec{k}$%
-value is a purely mathematical index which need not be identified with any
physical property of the system -- just as in (\ref{sumtransit}) the
physical meaning of the $k$-values is irrelevant to the mathematical
inequality. This observation indicated that it might well be possible to
extend the scope of the Mermin-Wagner theorem to partially restricted
systems as well.

Costache and Nenciu \cite{Cos Nen 1970} discuss the overall magnetization of
a partially finite three-dimensional Heisenberg model and reproduce the
generalization expected from the arguments by Chester et al. A similar,
though much more exhaustive, discussion has been given by Fisher and Jasnow
in two papers \cite{Fis Jas 1971a},\cite{Fis Jas 1971b}, where the authors
discuss Bose systems with respect to off-diagonal ordering, and spin systems
with respect to a magnetic phase transition. The general procedure is to
embed the space $\Omega $ occupied by the physical system into an enclosing
``box'' $\Lambda $ and, furthermore, to allow for a decomposition of $\Omega 
$ into ``subdomains'' $\Gamma .$ This makes it possible to distinguish
different boundary conditions -- such as a possible ``corridor'' surrounding 
$\Omega .$ This is of interest for the discussion of explicit bounds on e.g.
the spin-spin correlation function, but also on the limiting behaviour of
e.g. surface contributions in partially finite systems as the system
approaches the thermodynamic limit.\cite{Fis Jas 1971b} The ``global''
approach to partially restricted systems, i.e. the procedure of slicing up
physical space into the system's domain, its subdomains, and additional
auxiliary ``boxes'', however, appears to obscure the more basic question
whether or not within a given many-body system (e.g. of film geometry) a
phase transition can occur. This, as will be seen in the next section, is
quite independent of which numerical values certain model parameters display
in certain subdomains (apart from some general symmetry requirements.)

For the Heisenberg model, Corciovei and Costache \cite{Cor Cos 1967},\cite%
{Cor Cos 1972} have discussed the role of boundary conditions imposed on
partially finite systems, focussing on boundary conditions of the ``pinning
case''. The authors also discuss a classical analogue of the Bogoliubov
inequality (which, it appears, is derived independently of Mermin's paper %
\cite{Mer 1967}), which allows an alternative proof of the non-existence of
magnetization in the classical case. The case of the Hubbard model of thin
films is mentioned by Sukiennicki and Wojtczak \cite{Suk Woj 1972};
unfortunately no explicit proof of the absence of a phase transition is
given in their paper (contrary to what one of the authors suggests in ref. %
\cite{Suk 1976}).

\subsection{Systems with film geometries}

The problem of geometrical restriction imposed on physical systems, can be
made much more explicit, by defining the many-body system as a film system 
\textit{from the start.} Thus, no restrictions need to be imposed ``by
hindsight'', and any further specifications of interactions etc. can be
implemented on the level of the many-body model itself. A proof of the
Mermin-Wagner theorem for systems with film geometries has recently been
given \cite{Gel Nol 2000} for the Heisenberg, Hubbard, s-f and Kondo-lattice
models. In this section, we briefly outline the procedure, thereby extending
the validity of the Mermin-Wagner system to the Periodic Anderson Model with
film geometry (where we shall rule out a finite value of the layer
magnetization), and to a superconducting pairing mechanism in Hubbard films.

The film geometry is incorporated into the many-body Hamiltonian by
assigning each atom in the sample a double index $(n,\gamma ),$ where $n$
represents the vector $\vec{R}_{n}$ of the underlying 2D-Bravais lattice and
the Greek index $\gamma $ identifies which layer of the film is considered.
We shall assume that the film systems consists of $d$ identical layers
stacked on top of each other; the total number of lattice sites within one
layer is $N.$

For the \textbf{Periodic Anderson Model} the film Hamiltonian then is as
follows: 
\begin{eqnarray}
H_{PAM} &=&\sum_{ij\alpha \beta \sigma }t_{ij}^{\alpha \beta }c_{i\alpha
\sigma }^{\dagger }c_{j\beta \sigma }+\varepsilon _{f}\sum_{i\alpha \sigma
}f_{i\alpha \sigma }^{\dagger }f_{i\alpha \sigma }+\frac{U}{2}\sum_{i\alpha
}n_{i\alpha \uparrow }^{f}n_{i\alpha \downarrow }^{f}  \notag \\
&&+V\sum_{i\alpha \sigma }\left( c_{i\alpha \sigma }^{\dagger }f_{i\alpha
\sigma }+f_{i\alpha \sigma }^{\dagger }c_{i\alpha \sigma }\right)
-b\sum_{i\alpha }e^{-i\vec{K}\cdot \vec{R}_{i}}\left( \sigma _{c_{i\alpha
}}^{z}+\sigma _{f_{i\alpha }}^{z}\right)
\end{eqnarray}%
where the $c^{(\dagger )}$- and $f^{(\dagger )}$-operators denote fermionic
annihilation (destruction) operators for the electrons of the conduction
band and the $f$-electrons, respectively. (The number density operators $%
n_{i\alpha \sigma }^{f}=f_{i\alpha \sigma }^{\dagger }f_{i\alpha \sigma }$
and the $z$-components of the pseudo-spins, $\sigma _{c/f_{i\alpha
}}^{z}=(\hbar /2)\left( n_{i\alpha \uparrow }^{c/f}-n_{i\alpha \downarrow
}^{c/f}\right) ,$ are defined in the standard way.)

For the Bogoliubov inequality the operators $A$ and $C$ are chosen as 
\begin{eqnarray}
A_{(\gamma )} &=&\sigma _{c_{\gamma }}^{-}(-\vec{k}-\vec{K})+\sigma
_{f_{\gamma }}^{-}(-\vec{k}-\vec{K}) \\
C &\equiv &\sum_{\beta }C_{\beta }=\sum_{\beta }\left( \sigma _{c_{\beta
}}^{+}(\vec{k})+\sigma _{f_{\beta }}^{+}(\vec{k})\right) ,
\end{eqnarray}%
so that $A$ is \textit{layer-dependent,} whereas $C$ is \textit{%
layer-independent} (the sum $\sum_{\beta }$ extends over the whole sample.)
The (anti-)commutators appearing in the Bogoliubov inequality (\ref{bog})
can now be evaluated by making extensive use of the (anti-)commutation
relations for the conduction electrons and the $f$-electrons. The
Hamiltonian-independent quantities can be calculated in a straightforward
way:%
\begin{equation}
\left\langle \left[ C,A_{(\gamma )}\right] _{-}\right\rangle =\frac{2\hbar
^{2}N}{g_{J}\mu _{B}}M_{\gamma }(T,B_{0})
\end{equation}%
where the \textit{layer-dependent magnetization} $M_{\gamma }$ has been
introduced (again with a phase factor $e^{-i\vec{K}\cdot \vec{R}_{n}}$ to
account for possible antiferromagnetic ordering): 
\begin{equation}
M_{\gamma }(T,B_{0})=\frac{1}{N}\frac{g_{J}\mu _{B}}{\hbar }\sum_{n}e^{-i%
\vec{K}\cdot \vec{R}_{n}}\left\langle \sigma _{c_{n\gamma }}^{z}+\sigma
_{f_{n\gamma }}^{z}\right\rangle .
\end{equation}

The right-hand side of (\ref{bog}) can be bounded from above by%
\begin{equation}
\sum_{\vec{k}}\left\langle \left[ A,A^{\dagger }\right] _{+}\right\rangle
\leq 8\hbar ^{2}N^{2},
\end{equation}%
as can be shown by using a symmetrization procedure with subsequent
application of the spectral theorem, similar to the procedure outlined in
the appendix of \cite{Gel Nol 2000}.

For the Hamiltonian-dependent double commutator, one finds as an upper bound 
\begin{equation}
\left\langle \left[ \left[ C,H\right] _{-},C^{\dagger }\right]
_{-}\right\rangle \leq 4Nd\hbar ^{2}\left( \left| B_{0}M(T,B_{0})\right| +2%
\tilde{q}\vec{k}^{2}\right) ,
\end{equation}%
where the constant $\tilde{q}$ is taken to reflect the fact that the hopping
constants $t_{ij}^{\alpha \beta }$ fulfill the convergence criterion%
\begin{equation}
\frac{1}{Nd}\sum_{\gamma \beta }\sum_{nk}\left| t_{nk}^{\gamma \beta
}\right| \frac{(\vec{R}_{n}-\vec{R}_{k})^{2}}{4}\equiv \tilde{q}<\infty .
\end{equation}

Inserting these results into the ($k$-summed) Bogoliubov inequality, and
making the transition to the thermodynamic limit as in (\ref{transitiontotdl}%
), one arrives at an inequality for the layer magnetization $M_{\gamma }$
which is similar, though not identical with, the one in (\ref{magup}):%
\begin{equation}
M_{\gamma }^{2}(T,B_{0})\leq \xi \frac{\beta d}{\ln \left( 1+\frac{\xi
_{1}k_{0}^{2}}{\left| B_{0}M(T,B_{0})\right| }\right) },
\end{equation}%
the main difference being the factor $d$ (=number of layers) on the
right-hand side. Interestingly, the upper bound is proportional to the
inverse temperature $\beta $ and the number of layers $d.$ While for any
finite $\beta $ ($T\neq 0),$ and for any finite number of layers $(d<\infty
) $ a phase transition is ruled out because of the divergence of the
denominator as $B_{0}\rightarrow 0,$ the possibility of a phase transition
opens up if either $\beta $ or $d$ are infinite, i.e. when a two-dimensional
system is considered at $T=0$, or when the system becomes \textit{truly}
three-dimensional (i.e. infinitely extended in all dimensions).

As an example for a \textbf{superconducting phase transition,} we shall now
consider a pairing mechanism for \textit{Hubbard films. }Following the
discussion in \cite{Su Sch Zit 1997}, we restrict our attention to the order
parameter%
\begin{eqnarray}
\mathfrak{F}(\vec{K}) &=&\frac{1}{N\cdot d}\sum_{\alpha }\sum_{i}e^{-i\vec{K}%
\cdot \vec{R}_{i}}\left\langle c_{i\alpha \uparrow }^{\dagger }c_{i\alpha
\downarrow }^{\dagger }\right\rangle  \notag \\
&\equiv &\frac{1}{d}\sum_{i}\mathfrak{F}_{\alpha }(\vec{K})
\end{eqnarray}%
which measures the breakdown of U(1) symmetry due to local on-site pairing $(%
\mathfrak{F}_{\alpha }(\vec{K})$ is again the layer-dependent equivalent to
the bulk $\mathfrak{F}(\vec{K})).$ The ordering wave vector allows one to
distinguish between different types of pairing: for $\vec{K}=0$ one would
have $s$-wave pairing, for $\vec{K}\neq 0$ (generalized) $\eta $-pairing.
The Hubbard Hamiltonian with an appropriate U(1) symmetry-breaking
contribution of order $\lambda $ is then given by%
\begin{eqnarray}
H &=&\sum_{ij\alpha \beta \sigma }t_{ij}^{\alpha \beta }c_{i\alpha \sigma
}^{\dagger }c_{j\beta \sigma }+\frac{U}{2}\sum_{i\alpha \sigma }n_{i\alpha
\sigma }n_{i\alpha -\sigma }  \notag \\
&&-\lambda \sum_{i\alpha }\left( \eta _{i\alpha }^{+}e^{-i\vec{K}\cdot \vec{R%
}_{i}}+\eta _{i\alpha }^{-}e^{+i\vec{K}\cdot \vec{R}_{i}}\right)
\end{eqnarray}%
where we have introduced the operators%
\begin{eqnarray}
\eta _{i\alpha }^{+} &=&c_{i\alpha \uparrow }^{\dagger }c_{i\alpha
\downarrow }^{\dagger } \\
\eta _{i\alpha }^{-} &=&c_{i\alpha \downarrow }c_{i\alpha \uparrow } \\
\eta _{i\alpha }^{z} &=&\frac{1}{2}\left( n_{i\alpha \uparrow }+n_{i\alpha
\downarrow }-2\right) .
\end{eqnarray}

With the operators $A$ and $C$ chosen as%
\begin{eqnarray}
C &=&\sum_{\gamma }\eta _{\gamma }^{z}(\vec{k}) \\
A &=&\eta _{\alpha }^{+}(-\vec{k}-\vec{K}),
\end{eqnarray}%
the calculation of the (anti-)commutators for the Bogoliubov inequality
results in%
\begin{eqnarray}
\left\langle \left[ C,A\right] _{-}\right\rangle &=&\left\langle
\sum_{\gamma }\left[ \eta _{\gamma }^{z}(\vec{k}),\eta _{\alpha }^{+}(-\vec{k%
}-\vec{K})\right] _{-}\right\rangle =\left\langle \eta _{\alpha }^{+}(-\vec{K%
})\right\rangle \\
\sum_{\vec{k}}\left\langle \left[ A,A^{\dagger }\right] _{+}\right\rangle
&=&\sum_{\vec{k}}\sum_{ij}e^{i(\vec{k}+\vec{K})\cdot (\vec{R}_{i}-\vec{R}%
_{j})}\left\langle \left[ \eta _{\alpha j}^{+},\eta _{\alpha i}^{-}\right]
_{+}\right\rangle  \notag \\
&=&N\sum_{i}\left\langle \left[ \eta _{\alpha i}^{+},\eta _{\alpha i}^{-}%
\right] _{+}\right\rangle \leq 4N^{2} \\
\left\langle \left[ \left[ C,H\right] _{-},C^{\dagger }\right]
_{-}\right\rangle &\leq &Nd\tilde{q}\vec{k}^{2}+2\lambda Nd\left| \mathfrak{F%
}(\vec{K})\right| .
\end{eqnarray}%
Thus, in this case as well, one gets a result of the form 
\begin{equation}
\frac{v_{0}^{(2)}}{2\pi }\left| \mathfrak{F}_{\alpha }(\vec{K})\right|
^{2}\leq \frac{\beta d}{\ln \left( 1+\frac{\tilde{q}k_{0}^{2}}{2\lambda
\left| \mathfrak{F}(\vec{K})\right| ^{2}}\right) }
\end{equation}%
which excludes a finite value of the layer-specific quantity $\mathfrak{F}%
_{\alpha }$, again due to the divergence of the denominator in the limit $%
\lambda \rightarrow 0$. Thus, no pairing transition of the proposed kind and
hence \textit{no corresponding superconducting phase transition can occur in
Hubbard films,} provided the number of layers is finite and the temperature
is non-zero.

The dependence on $d$ of the upper bound in the film cases, confirms a
statement by Fisher \cite{Fis 1973}, which is based on his work with Jasnow %
\cite{Fis Jas 1971b}, where the upper bound scales with the extension of the
``box'' $\Lambda $ in which the physical system is embedded. Note, however,
that the two approaches start from opposite ends: Fisher and Jasnow impose
external geometrical constraints, whereas the approach presented here (and
the related approach of ref. \cite{Gel Nol 2000}) includes the film geometry
in the Hamiltonian from the very beginning.

\subsection{Fractal lattices}

The discussion in the previous section points to a deeper connection between
the geometry of the sample and the interaction between its constituents.
Hattori et al. \cite{Hat Hat Wat 1987} have pointed to an interesting aspect
of this relation, and we shall briefly summarize their discussion.

In the theory of phase transitions, and particularly in the versions of the
Mermin-Wagner theorem discussed so far, one typically deals with a spin
system on a translationally invariant lattice which is embedded into a
Euclidean space of integer dimension $(D=1,2,$ or $3)$. It turns out that
the conditions for the (non-)existence of a phase transition are governed by
the dimension of this Euclidean space. More generally, the critical
properties and the scaling limits of such a many-particle system crucially
depend on the Euclidean metric associated with the dimensionality:
Correlation functions of the form $\left\langle \phi _{n}...\phi
_{m}\right\rangle $ are Euclidean invariant functions in the scaling limit.
This, however, poses a question which Hattori et al. put forward in the
following way: ``Why do the spins on the lattice `know' this natural
embedding into Euclidean space that should govern their critical
phenomena?'' After all, the only geometrical structure that spins can feel
is the one given by the Hamiltonian, i.e. by the ``network structure of the
interaction (the kinetic term, in terms of field theory).''\cite{Hat Hat Wat
1987}

As long as one is concerned with highly regular, i.e. translationally
invariant lattices, this duality between the structure of the interaction
(as contained in the Hamiltonian) and Euclidean dimension remains somewhat
hidden. One instance where it surfaces, however, is the complementarity,
mentioned in the last paragraph of the previous section, between Fisher and
Jasnow's approach (which suggests that the upper bound in the Mermin-Wagner
approach is $\sim \Lambda ,$ i.e. proportional to the linear extension of
the ``\textit{embedding} box'') and our calculation which proves that the
upper bound in a film system is indeed proportional to the \textit{number of
layers }$d$ (by virtue of the structure of the \textit{Hamiltonian}).

The question becomes highly relevant for irregular lattices, where many of
the techniques (e.g. Fourier transforms, or notions of ``momentum'' vectors)
cannot be applied any longer. As a step towards discussing such irregular
systems, Hattori et al. develop a Gaussian field theory on a general
network; the authors succeed in giving a definition of the spectral
dimension of general networks in terms of the critical behaviour of a spin
system, and show how, for a restricted class of networks, the spectral
dimension can be determined in practice. While a detailed discussion of
their results would be far beyond the scope of this paper, an application by
Cassi \cite{Cas 1992} is of interest. By applying Hattori et al.'s results
to the case of fractal and disordered lattices whose spectral dimension is
less or equal to 2, it is shown that at finite temperature no spontaneous
magnetization can exist for the classical $O(n)$ and the ferromagnetic
(quantum) Heisenberg model. This is due to the infrared divergencies
displayed by the corresponding Gaussian models and confirms once again the
wide applicability of the Mermin-Wagner theorem.

\section{On the validity of the Mermin-Wagner theorem in approximative
theories}

Surprisingly few authors have commented on the relevance of the
Mermin-Wagner theorem for methods of approximation, some of which
consistently predict finite-temperature phase transitions in one or two
dimensions, even when applied to many-body models that obey the
Mermin-Wagner theorem. Sometimes the breaking of the Mermin-Wagner theorem
has been taken to be an advantage; Sukiennicki and Wojtczak, for example,
believe ``that the molecular field approximation of the Hubbard model
approaches the physical reality (magnetic order in thin films does really
exist) better than the Hubbard model in its exact form.''\cite{Suk Woj 1972}
While this may seem \textit{prima facie} reasonable, it is difficult to
maintain this view in the light of the advances that have been achieved
within the (unmodified) Hubbard model by disposing with simple molecular
field approximations and adopting instead more refined methods of
approximation. If one aspires to formulate criteria for the reliability of
one's methods of approximation (other than justifying them in an \textit{ad
hoc} fashion), one will need to shed some light on the relation between the
underlying many-body model, the exact results known about it, and the
mechanism of approximation. It would seem unsatisfactory to rely blindly on
the method of approximation introducing exactly the kind of
symmetry-breaking needed to reflect physical reality. If it turns out that
certain classes of approximation methods consistently produce reasonable
results for quasi-2D cases, this needs explanation. One could even turn the
argument around and question methods of approximation for the 3D case in
case they violate the Mermin-Wagner result in two dimensions. As Walker puts
it: ``[A]pproximation schemes that have been applied to real solids can
equally well be applied to one- and two-dimensional solids. If these
approximation schemes predict the occurrence of spontaneous magnetization in
one and two dimensions as well as in three dimensions [...], the validity of
these predictions in three dimensions should clearly be investigated more
fully.''\cite{Wal Rui 1968}

Considerations of this sort set a possible agenda for future research into
the relevance of exact results for the application of approximation schemes.
In the rest of this section, we shall present a fairly elementary discussion
of the Heisenberg model within different methods of approximation. In
particular, we shall contrast two possible ways of testing ``Mermin-Wagner
type'' behaviour.

Consider the spin-1/2 anisotropic Heisenberg model given by 
\begin{equation}
H=-\sum_{n.n.ij}\left( J_{ij}\left(
S_{i}^{x}S_{j}^{y}+S_{i}^{y}S_{j}^{y}\right)
+J_{ij}^{z}S_{i}^{z}S_{j}^{z}\right)  \label{anisHei}
\end{equation}%
where $J_{ij}^{z}=\varepsilon J_{ij}>J_{ij}$ and the summation index $%
``n.n." $ refers to summation over nearest neighbours only. This kind of
anisotropy in spin-space favouring uniaxial ordering is widely used and
should, even in 2D, break the symmetry of the corresponding isotropic model.

It has sometimes been claimed \cite{Tus Szc Log 1995} that a modification of
mean-field theory, known as Onsager reaction-field theory, can reproduce the
Mermin-Wagner theorem in the case of the Heisenberg model. According to the
Onsager theory the orienting part of the mean field acting on a given spin
must not include that part of the contribution from the spins in the
vicinity which is due to their correlation with the given spin under
consideration. Thus, one must stipulate that the full mean field decomposes
into two independent contributions, the correlated and the uncorrelated one,
called \textit{reaction field} and \textit{cavity field,} respectively. (For
a detailed discussion see \cite{Gel 2001}.) Singh \cite{Sin 2000} has
succeeded in applying Onsager's reaction-field idea to the anisotropic
Heisenberg model as given by (\ref{anisHei}). It turns out that the Onsager
reaction-field result for the critical temperature, $T_{c}^{Ons},$ is
related to that of ordinary mean-field (MF) theory by the equation%
\begin{equation}
\frac{T_{c}^{Ons}}{T_{c}^{MF}}=\frac{1}{\sum\limits_{\vec{k}}\frac{1}{\left(
1-\frac{J^{z}(\vec{k})}{J^{z}(\vec{K})}\right) }}  \label{onsager}
\end{equation}%
where $\vec{K}$ denotes the ordering wave vector and $J^{z}(\vec{k})$ is the
standard Fourier transform of the exchange integrals $J_{ij}$. From (\ref%
{onsager}), however, we see that while Onsager reaction-field theory does
modify the ordinary mean-field result, it is independent of the anisotropy
assumed above by setting $J_{ij}^{z}=\varepsilon J_{ij},$ since the
anisotropy parameter $\varepsilon $ will drop out of the fraction $J^{z}(%
\vec{k})/J^{z}(\vec{K})$ in the denominator. Effects of dimensionality that
stem from the summation $\sum_{\vec{q}}$ are, on the other hand, retained,
such as the denominator's logarithmic divergence in two dimensions, which
rules out a phase transition. This compliance with the Mermin-Wagner
theorem, however, is rendered spurious by the fact that no anisotropy
effects are captured by the Onsager approach.

It is worthwhile to explore whether more sophisticated methods of
approximation are more successful in accounting for effects due to (reduced)
dimensionality and (possibly anisotropic) interaction. For the Tyablikov
procedure of decoupling higher-order Green's functions within the equation
of motion scheme, this is indeed the case. As has been shown in \cite{Gel
2001}, the Tyablikov procedure accords with the Mermin-Wagner theorem in two
dimensions: Only when an external field is applied, can there be a finite
value of the magnetization at non-zero temperature. The proof is based on a
series expansion of the (implicit) equation for the expectation value $%
\left\langle S^{z}\right\rangle $, 
\begin{equation}
\frac{\left\langle S^{z}\right\rangle }{\hbar S}=1/\left( \frac{1}{N}\sum_{%
\vec{k}}\coth \left( \frac{\beta }{2}E(\vec{k})\right) \right) ,
\label{implicsz}
\end{equation}%
where the energy $E(\vec{k})$ is given by $E(\vec{k})=2\hbar \left\langle
S^{z}\right\rangle \left( J_{0}-J(\vec{k})\right) +g_{J}\mu _{B}B_{0}.$ As a
result of the Tyablikov approximation, (\ref{implicsz}) is taken to be
uniform for all spins in the lattice, which seems to limit the discussion to
the ferromagnetic case. However, it has also been shown that an analogous
result holds for the case of spontaneous sublattice magnetization in
ABAB-type antiferromagnets. The corresponding formula for the N\'{e}el
temperature in that case is of the following form: 
\begin{equation}
T_{N}^{Tyab}=\frac{J_{0}}{2k_{B}}\frac{1}{\frac{2}{N}\sum\limits_{\vec{k}}%
\frac{1}{\left( 1-\frac{J^{2}(\vec{k})}{J_{0}^{2}}\right) }}.
\end{equation}%
It is straightforward to show that in the thermodynamic limit the sum in the
denominator diverges in two dimensions while remaining finite in three
dimensions. What remains to be shown is that the Tyablikov method can make
sense of anisotropy effects, i.e. perform better than the Onsager
reaction-field theory in accounting for ordering due to anisotropy. One such
test is whether or not the method can correctly predict the Mermin-Wagner
theorem if one starts from the three-dimensional case and gradually turns
off interlayer coupling. If we choose the interaction parameters $J_{ij}$ so
that (for a fixed lattice site $i)$ $J_{ij}=J_{||}$ if sites $i$ and $j$ lie
within a plane, and $J_{ij}=J_{\perp }$ if $j$ is located in the direction
orthogonal to this plane, then, by varying the anisotropy parameter%
\begin{equation}
\varepsilon :=\frac{J_{\perp }}{J_{||}}
\end{equation}%
we can change the coupling continuously from quasi two-dimensional $%
(\varepsilon \equiv 0)$ to (isotropic) three-dimensional $(\varepsilon =1).$
One can then, indeed, prove that in the quasi-2D limit of vanishing
interlayer coupling $(\varepsilon \rightarrow 0)$ the N\'{e}el temperature
(as calculated from the Tyablikov method) vanishes as 
\begin{equation}
T_{N}^{Tyab}%
\begin{array}{c}
\\ 
\widetilde{\varepsilon \rightarrow 0}%
\end{array}%
\frac{\varepsilon }{\left| \ln (\varepsilon )\right| },
\end{equation}%
in accordance with the Mermin-Wagner theorem. In its original form, the
Tyablikov method applies to the $S=1/2$ case only, and so, by definition, do
the results derived so far. An extension of the results to higher spins can,
however, be given in a straightforward way \cite{Gel 2001}.

In this section, we have contrasted two approximation schemes, the Onsager
reaction-field theory and the Tyablikov method. The failure of
reaction-field theory to capture the effect of anisotropic coupling suggests
that it does not adequately capture the behaviour of the Heisenberg model as
far as the existence, or absence, of a phase transition is concerned. The
Tyablikov method, despite its still rather simple random-phase
characteristics, seems to fare better: It can be analytically shown to
produce all the relevant limiting cases of possible ``Mermin-Wagner type
behaviour,'' and it also correctly reproduces the qualitative change
associated with the transition from three to two dimensions by introducing
anisotropic (and eventually vanishing) interlayer coupling.

\section{Summary}

In this paper, we have presented a survey of recent and established results
concerning the application of Bogoliubov's inequality to the theory of phase
transitions. This includes the classic papers by Hohenberg, Mermin, and
Wagner as well as a host of other proofs concerning a variety of different
many-body models, order parameters and system geometries. In particular, new
proofs for the absence of superconducting long-range order in Hubbard films,
and of magnetic long-range order in Periodic Anderson Model films have been
presented. The complementarity between geometric constraints on the one hand
and anisotropy on the other hand, has been sketched in theoretical terms,
and has been applied to two specific methods of approximation, Onsager
reaction-field theory and the Tyablikov method. The two methods discussed
here, can only give an indication as to which effects and limiting cases
should be considered, if one attempts to characterize a method in terms of
its accordance with the Mermin-Wagner theorem. The more general problem of
the relevance of exact results for the numerical or approximative treatment
of many-body models, as characterized in section 4, will almost certainly
remain and in our view deserves closer attention.


\begin{thebibliography}{99}
\bibitem{Bog 1962} Bogoliubov, N.N.: Phys. Abhandl. Sowjetunion \textbf{6}
(1962), 1, 113, 229

\bibitem{Bog 1960} Bogoliubov, N.N.: Physica (Suppl.) \textbf{62} (1960) S1

\bibitem{Nol 1986} Nolting, W.: \textit{Quantentheorie des Magnetismus,}
vol. 2, Stuttgart: B.G. Teubner 1986

\bibitem{Fro Pfi 1981} Fr\"{o}hlich, J. and C. Pfister: Commun. Math. Phys. 
\textbf{81} (1981) 277-298

\bibitem{Lut 1966} Luttinger, J.M.: Progr. Theor. Phys. Suppl. (Kyoto) 
\textbf{37/38} (1966) 35

\bibitem{Wag 1966} Wagner, H.: Z. Phys. \textbf{195} (1966) 273

\bibitem{Kub Kis 1990} Kubo, K. and T. Kishi: Phys. Rev. B \textbf{41}
(1990) 4866

\bibitem{Pit Str 1991} Pitaevskii, L. and S. Stringari: J. Low Temp. Phys. 
\textbf{85} (1991) 377

\bibitem{Hoh 1967} Hohenberg, P.C.: Phys. Rev. \textbf{158} (1967) 383-386

\bibitem{Mer Wag 1966} Mermin, N.D. and H.\ Wagner: Phys. Rev. Lett. \textbf{%
17} (1966) 1133-1136

\bibitem{Gel Nol 2000} Gelfert, A. and W. Nolting: phys. stat. sol. (b) 
\textbf{217} (2000) 805-818

\bibitem{Mer 1967} Mermin, N.D.: J. Math. Phys. \textbf{8} (1967) 1061

\bibitem{Wal 1967} Walker, M.B.: Can. J. Phys. \textbf{46} (1968) 817-821

\bibitem{Mer 1968} Mermin, N.D.: Phys. Rev. \textbf{176} (1968) 250-254

\bibitem{Gri 1972} Griffiths, R.B.: ``Rigorous Results and Theorems,'' in
Domb, C. and M.E. Green (eds.): \textit{Phase Transitions and Critical
Phenomena, }New York: Academic Press 1972, 8-109

\bibitem{Fer 1970b} Fern\'{a}ndez, J.F.: Phys. Rev. A \textbf{2} (1970)
2555-2559

\bibitem{Kis She 1972} Kishore, R. and D. Sherrington: Phys. Lett. A \textbf{%
42} (1972) 205-207

\bibitem{Sch 1977} Schuster, H.G.: Phys. Lett. A \textbf{60} (1977) 89-91

\bibitem{Gul Gul 1996} Gul\'{a}sci, Z. and M. Gul\'{a}sci: Advances in
Physics \textbf{47} (1981) 1-89

\bibitem{Bra Hol 1994} Bramwell, S.T. and P.C.W. Holdsworth: Phys. Rev. B 
\textbf{49} (1994) 8811-8814

\bibitem{Weg 1967} Wegner, F.: Phys. Lett. A \textbf{24} (1967) 131-132

\bibitem{Wal Rui 1968} Walker, M.B. and T.W. Ruijgrok: Phys. Rev. \textbf{171%
} (1968) 513-515

\bibitem{Lie Mat 1962} Lieb, E. and D. Mattis: Phys. Rev. \textbf{125}
(1962) 164

\bibitem{Gho 1971} Ghosh, D.K.: Phys. Rev. Lett. \textbf{27} (1971) 1584-1587

\bibitem{Ber Ver 1974} van den Bergh, M. and G. Vertogen: Phys. Lett. A 
\textbf{50} (1974) 85-87

\bibitem{Rob Mic 1976} Robaszkiewicz, S. and R. Micnas: phys. stat. sol. (b) 
\textbf{73} (1976) K35-K39

\bibitem{Bar Yab 1975} Baryakhtar, V.G. and D.A. Yablonskii: phys. stat.
sol. (b) \textbf{72} (1975) K95-K98

\bibitem{Tho 1971} Thorpe, M.F.: J. Appl. Phys. \textbf{42} (1971) 1410-1411

\bibitem{Krz 1976} Krzeminski, S.: phys. stat. sol. (b) \textbf{74} (1976)
K119-K123

\bibitem{Rit Mav 1978} Ritchie, D.S. and C. Mavroyannis: J. Low Temp. Phys. 
\textbf{32} (1978) 813-820

\bibitem{Uhr 1992} Uhrig, G.S.: Phys. Rev. B \textbf{45} (1992) 4738-4740

\bibitem{Mat Mat 1997} Matayoshi, G. and S. Matayoshi: Bull.\ Coll. Sci.
Univ. Ryukyus \textbf{64} (1997) 1-5

\bibitem{Ras Tas 1989} Rastelli, E. and A. Tassi: Phys.\ Rev. B \textbf{40}
(1989) 5282-5284

\bibitem{Pro Lop 1983} Proetto, C. and A. Lopez: J. Physique - Lettr. 
\textbf{44} (1983) L635-L640

\bibitem{Noc Cuo 1999} Noce, C. and M. Cuoco: Phys. Rev. B \textbf{59}
(1999) 7409-7412

\bibitem{Su Sch Zit 1997} Su, G.; Schadschneider, A.; and J. Zittartz: Phys.
Lett. A \textbf{230} (1997) 99-104

\bibitem{Che Fis Mer 1969} Chester, G.V.; Fisher, M.E.; and N.D. Mermin:
Phys.\ Rev. \textbf{185} (1969) 760-761

\bibitem{Cos Nen 1970} Costache, G. and G. Nenciu: Phys. Lett. A \textbf{33}
(1970) 193-194

\bibitem{Fis Jas 1971a} Fisher, M.E. and D. Jasnow: Phys. Rev. B \textbf{3}
(1971) 895-907

\bibitem{Fis Jas 1971b} Fisher, M.E: and D. Jasnow: Phys. Rev. B \textbf{3 }%
(1971) 907-924

\bibitem{Cor Cos 1967} Corciovei, A. and G. Costache: Phys. Lett. A \textbf{%
25} (1967) 458-459

\bibitem{Cor Cos 1972} Corciovei, A. and G. Costache: Rev. Roum. Phys. 
\textbf{17} (1972) 541-546

\bibitem{Suk Woj 1972} Sukiennicki, A. and L. Wojtczak: Phys. Lett. A 
\textbf{41} (1972) 37-38

\bibitem{Suk 1976} Sukiennicki, A.: IEEE Trans. Magn. \textbf{12} (1976)
90-95

\bibitem{Fis 1973} Fisher, M.E.: J.\ Vac. Sci. Technol. \textbf{10}
(1973)665-673

\bibitem{Hat Hat Wat 1987} Hattori, K.; Hattori, T.; and H. Watanabe: Progr.
Theor. Phys. (Suppl.) \textbf{92} (1987) 108-143

\bibitem{Cas 1992} Cassi, D.: Physica A \textbf{191} (1992) 549-553

\bibitem{Tus Szc Log 1995} Tusch, M.A.; Szchech, Y.H.; and D.E. Logan: Phys.
Rev. B \textbf{53} (1995) 5505

\bibitem{Gel 2001} Gelfert, A.: ``\textit{On the role of dimensionality in
many-body theories of magnetic long-range order,'' }Los Alamos National
Laboratory electronic archives: arXiv:cond-mat/0106031 , 100 pp.

\bibitem{Sin 2000} Singh, A.: ``\textit{Breakdown of the Onsager reaction
field theory in two dimensions,'' }Los Alamos National Laboratory electronic
archives: arXiv:cond-mat/0003040 , 3 pp.
\end{thebibliography}
\end{document}